\documentclass[11pt]{article}
\usepackage[left=1in,top=1in,right=1in,bottom=1in,head=.1in]{geometry}
\usepackage{amsmath}
\usepackage{color}
\usepackage{multirow}
\usepackage{graphicx}
\usepackage{amssymb, amsthm}
\usepackage{url}
\usepackage{setspace}
\usepackage{times}
\usepackage{subcaption}
\usepackage{float}
\usepackage{bbm}

\RequirePackage{amsmath,amssymb}
\RequirePackage{amsthm}
\RequirePackage{natbib}
\RequirePackage[colorlinks,allcolors=blue]{hyperref}

\theoremstyle{definition}

\newtheorem{assumption}{Assumption}

\newcommand\indep{\protect\mathpalette{\protect\independenT}{\perp}}
\def\independenT#1#2{\mathrel{\rlap{$#1#2$}\mkern2mu{#1#2}}}
\newcommand{\E}{\ensuremath{\mathbb{E}}}

\newcommand{\R}{\ensuremath{\mathbb{R}}}

\newcommand{\calM}{\ensuremath{\mathcal{M}}}

\newcommand{\calT}{\ensuremath{\mathcal{T}}}

\newcommand{\calN}{\ensuremath{\mathcal{N}}}

\newcommand{\calH}{\ensuremath{\mathcal{H}}}

\newcommand{\simiid}{\overset{\textrm{i.i.d.}}{\sim}}

\newcommand{\<}{\langle}
\renewcommand{\>}{\rangle}

\title{Is it who you are or where you are? Accounting for compositional differences in cross-site treatment variation
\thanks{This research was supported in part by the Institute of Education Sciences, U.S. Department of Education, through Grant R305D200010. The opinions expressed are those of the authors and do not represent views of the Institute or the U.S. Department of Education. This material is based upon work supported by the National Science Foundation under Grant No. 1745640.}}
\author{Benjamin Lu, Eli Ben-Michael, Avi Feller, and Luke Miratrix\\[1em]UC Berkeley and Harvard University}
\date{\today}
\begin{document}
\maketitle

\begin{abstract}
Multisite trials, in which treatment is randomized separately in multiple sites, offer a unique opportunity to disentangle treatment effect variation due to ``compositional'' differences in the distributions of unit-level features from variation due to ``contextual'' differences in site-level features. 
In particular, if we can re-weight (or ``transport'') each site to have a common distribution of unit-level covariates, the remaining effect variation captures contextual differences across sites. In this paper, we develop a framework for transporting effects in multisite trials using approximate balancing weights, where the weights are chosen to directly optimize unit-level covariate balance between each site and the target distribution. We first develop our approach for the general setting of transporting the effect of a single-site trial. We then extend our method to multisite trials, assess its performance via simulation, and use it to analyze a series of multisite trials of welfare-to-work programs. Our method is available in the {\ttfamily balancer} {\ttfamily R} package.
\end{abstract}

\clearpage
\onehalfspacing
\section{Introduction}
\label{sec:intro}
Multisite randomized controlled trials (RCTs) randomize treatment assignment separately within each of several sites. Although they can be more complicated to plan and execute than single-site RCTs \citep{kraemer2000}, they offer several methodological advantages \citep{weinberger2001} and have been conducted across a variety of domains. Multisite RCTs have been used to study, for example, the effects of the Head Start program on childhood educational outcomes \citep{puma2010}, of welfare-to-work programs on participant earnings \citep{riccio1992, kemple1994}, of police body-worn camera usage on citizen complaints \citep{ariel2017}, and of psychoeducational interventions on cancer patients' emotional health \citep{stanton2005}.

Multisite RCTs show promise in part because they can reveal how  treatment effects vary across different settings. Specifically, they can help disentangle treatment effect variation due to ``compositional'' differences in sites' distributions of unit-level features from variation due to ``contextual'' differences in site-level features \citep{rudolph2018}. For example, a multisite RCT might allow researchers to determine how much of the variation in an educational intervention's effects across schools is due to differences in the type of school (\textit{e.g.}, whether it is public, private, or charter) as opposed to differences in student population characteristics (\textit{e.g.}, family income, parents' level of education, and race). This type of decomposition can then inform decisions about where and when to implement the intervention. Statistically, this analysis can be framed as a special case of a broad class of problems seeking to ``transport'' or ``generalize'' treatment effects from a given site to a target population. In particular, if we transport the treatment effect from each site to the same target distribution of unit-level covariates, we can attribute any remaining variation in the transported treatment effects to differences in site-level features and unobserved unit-level features.

Many transportation and generalization methods rely on weighting estimators, including doubly robust estimators that combine weighting and outcome modeling \citep[see][]{egamielements2021}. Traditional inverse propensity score weighting (IPW) is the workhorse method \citep[\textit{e.g.},][]{rudolph2018}
but can perform poorly with many covariates or with extreme estimated propensity scores.
IPW is also limited by requiring unit-level data in the target distribution. More recently, \citet{josey2020} proposed to instead use entropy balancing \citep{hainmueller2012}, which seeks to find weights that exactly balance a few covariates. But, while promising, entropy balancing is often infeasible with even a moderate number of covariates.

In this paper, we develop a framework for transporting treatment effects using approximate balancing weights, a method recently developed in the observational causal inference literature \citep{Zubizarreta2015, hirshberg2020, benmichael2021review}. Our approach chooses the weights to directly optimize unit-level covariate balance between each site and the target distribution. Unlike existing estimators, this approach can accommodate high-dimensional covariates, including higher-order interactions and kernels, and can target an arbitrary covariate distribution without requiring unit-level data. We first develop our approach for the general task of transporting the treatment effect from a single site. We then adapt it to our motivating special case of decomposing treatment effect variation in multisite trials. To decompose treatment effect variation, we first transport every site's treatment effect to a common target distribution of unit-level covariates. We can then descriptively analyze the variation in these transported treatment effects (net of unit-level differences), including their relationship to site-level covariates. 

\begin{figure}
    \centering
    \includegraphics[width = 0.99\linewidth]{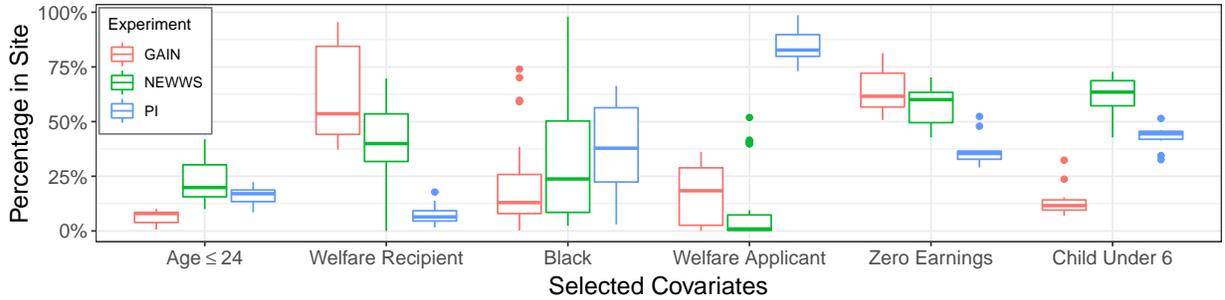}
    \caption{Distributions of the marginal prevalence of selected binary covariates across the 59 sites of the multisite welfare-to-work experiments analyzed in this paper. The covariates are distributed differently across sites. For example, each site in Project Independence had a greater proportion of welfare applicants than any site in the other two multisite experiments.  }
    \label{fig:imbal_demo}
\end{figure}

We apply our approach to a collection of seminal studies on welfare-to-work policies between 1988 and 1994. Specifically, we re-examine three separate multisite experiments: Project GAIN, Project Independence (PI), and the National Evaluation of Welfare-to-Work Strategies (NEWWS). These multisite experiments collectively spanned 59 sites across seven states, totaling 69,399 participants. Within each site, participants were randomly assigned either to attend or to be barred from a job training program. The primary outcomes were employment status and earnings two years after random assignment. For each participant, we also observe 23 pre-treatment covariates, including earnings prior to randomization, number of dependent children, and high school completion. As Figure \ref{fig:imbal_demo} shows, these covariates are distributed differently in each site and across the three experiments as a whole. As a result, direct comparisons of treatment effects across individual sites and the overall experiments might be difficult to interpret. 

We therefore use our method to transport the treatment effects from the 59 sites to the same target covariate distribution. After re-weighting, however, we find that differences in sites' observed unit-level covariate distributions explain little of the treatment effect heterogeneity between sites---about 10 percent, by one measure we use. Other factors, like differences in site-level features, likely play a greater role.

The remainder of this paper is organized as follows. 
Section \ref{sec:setup} establishes the basic setting for our problem, identifies our estimand, and distinguishes our work from recent literature. 
Section \ref{sec:estimation} introduces our proposed estimator based on approximate balancing weights. 
Section \ref{sec:multi} shows how our setup and proposed estimators naturally extend to multisite RCTs and sketches our framework for decomposing treatment effect variation in this context. 
Section \ref{sec:sim} compares our proposed estimator to other standard estimators via simulation. 
Section \ref{sec:app} applies our method to investigate treatment effect variation in the welfare-to-work experiments. 
Section \ref{sec:conc} concludes.

\section{Setup}
\label{sec:setup}
\subsection{Estimand and Assumptions}
\label{sec:assumptions}

We first consider a single-site RCT with a binary treatment, turning to multisite RCTs in Section \ref{sec:multi}. 
Of the $n$ units in the experiment, $n_1$ are assigned to treatment and $n_0$ are assigned to control. Let $Z_i \in \{0, 1\}$ be a binary treatment indicator for the $i^{\text{th}}$ unit. In addition to treatment assignment, we observe for each unit a vector of pre-treatment covariates $X_i$ and the unit's post-treatment outcome $Y_i$.
We adopt the potential outcomes framework \citep{neyman1923, rubin1974} and assume that each unit has a potential outcome under treatment $Y_i(1)$ and a potential outcome under control $Y_i(0)$; implicit in this notation is the stable unit treatment value assumption. We also assume stability of potential outcomes, so that $Y_i = Z_iY_i(1) + (1 - Z_i)Y_i(0)$. Additionally, we assume that the units in the experiment are drawn i.i.d. from some distribution $P$: $\{X_i, Z_i, Y_i(0), Y_i(1)\} \simiid P$ for $i = 1, \ldots, n$.

Throughout this paper, we use several conditional expectations defined over the experimental population's distribution $P$. First we denote the propensity score for the experimental population as $e(x) := \E_{P}\left(Z \mid X = x\right)$. To simplify notation, we assume throughout the main text that this is some constant $\pi$ known by design. But our results are easily generalized to non-constant propensity scores. Second, we denote the prognostic score, the expected control potential outcome conditioned on the observed covariates, as $m_0(x) := \E_P\{Y(0) \mid X = x\}$. Analogously, we denote the expected potential outcome under treatment conditioned on the observed covariates as $m_1(x) := \E_{P}\left\{Y(1) \mid X = x\right\}$. Finally, we denote the conditional average treatment effect (CATE) function as 
\[ \tau(x) := \E_{P}\left\{Y(1) - Y(0) \mid X = x\right\} . \]

We seek to characterize the average treatment effect (ATE) for some target population beyond the finite population of experimental units at hand. Let $P^\ast$ denote the distribution of this target population. The experimental population can but need not be a subset of the target population. Our primary estimand is the ATE in the experimental population if it had the same distribution of observed covariates as the target population:
\begin{equation}
\label{eq:estimand}
    \tau^\ast := \E_{P^\ast}\left[\E_{P}\left\{Y(1) - Y(0) \mid X\right\}\right] = \E_{P^\ast}\left\{\tau(X)\right\} = \int \tau(x) dP^\ast(x).
\end{equation}
The inner expectation in \eqref{eq:estimand} is taken over the \textit{experimental} population's distribution of potential outcomes conditioned on $X$; the outer expectation in \eqref{eq:estimand} is taken over the \textit{target} population's distribution of observed covariates $X$.

Because $\tau^\ast$ represents an effect hybridized between the target and experimental populations, it can provide a useful framing for decomposing treatment effect variation across different populations. Heuristically, because the CATE function in both $\tau^\ast$ and the experimental population's ATE is the same,
\begin{equation}
\tau := \E_{P}\left\{Y(1) - Y(0)\right\} = \E_{P}\left[E_{P}\left\{Y(1) - Y(0) \mid X\right\}\right] = \E_{P}\left\{\tau(X)\right\} = \int\tau(x) dP(x)
\end{equation}
we can attribute any difference between $\tau$ and $\tau^\ast$ to differences in the observed covariate distributions of the target and experimental populations. And, as we show in Section \ref{sec:multi} when we extend our setup to multisite RCTs, analogous reasoning in the opposite direction---comparing ATEs defined by different CATE functions integrated over the same distribution of observed unit-level covariates---can help isolate treatment effect variation due to site-level or unobserved unit-level covariate differences between sites. 

In addition to the foundational assumptions stated at the start of this section, we make three assumptions that together identify $\tau^\ast$. We assume that treatment is randomized, that the probability of treatment is strictly between 0 and 1, and that the target distribution of observed covariates overlaps with the experimental distribution of observed covariates.
\begin{assumption}[Randomization]
\label{as:random}
Each unit's treatment assignment is independent of its potential outcomes: $Y(0), Y(1) \indep Z$ for $\{X, Z, Y(0), Y(1)\} \sim P$.
\end{assumption}
\begin{assumption}[Positivity of Treatment Assignment]
\label{as:positivity}
If $x$ satisfies $p(x) > 0$, then $0 < e(x) < 1$.
\end{assumption}
\begin{assumption}[Absolute Continuity]
\label{as:cont}
The target distribution of observed covariates is absolutely continuous with respect to the experimental distribution of observed covariates: $p^\ast(x) \ll p(x)$.
\end{assumption}
\noindent One advantage of the framework we have established thus far is that the assumptions involved are fairly light. Randomization and positivity of treatment assignment are typically satisfied by design in RCTs. And the target population can be chosen to satisfy absolute continuity, although this can be more difficult in high dimensions. (See, for example, \citet{d2017} for discussion of issues that can arise.) Under Assumptions \ref{as:random}-\ref{as:cont}, $\tau^\ast$ can be identified by
\begin{equation}
\label{eq:identify}
\begin{aligned}
    \tau^\ast &= \E_{P^\ast}\left\{\E_{P}\left(Y \mid Z = 1, X\right) - \E_{P}\left(Y \mid Z = 0, X\right)\right\}\\
    &= \E_{P}\left[\left\{\frac{ZY}{e(X)} - \frac{(1 - Z)Y}{1 - e(X)}\right\} \frac{dP^\ast}{dP}\right].
\end{aligned}
\end{equation}

Throughout this paper, we rely only on Assumptions \ref{as:random}-\ref{as:cont} or, in Section \ref{sec:multi}, their extensions to multisite RCTs. However, we pause here to note that other work often invokes some form of the following assumption as well \citep{dahabreh2018}.
\begin{assumption}[Mean Generalizability of Treatment Effects]
\label{as:general}
For all $x$ satisfying $p^\ast(x) > 0$,
\[
\E_{P^\ast}\left\{Y(1) - Y(0) \mid X = x\right\} = \E_{P}\left\{Y(1) - Y(0) \mid X = x\right\}.
\]
\end{assumption}
\noindent Assumption \ref{as:general} implies that the CATE function is the same in both the experimental population and the target population wherever the target covariate distribution has positive density. A common variation of this assumption is that units' potential outcomes are independent of the population to which they belong, conditional on covariates \citep{hotz2005, flores2013, allcott2015}. Under Assumption \ref{as:general}, our estimand $\tau^\ast$ is equivalent to the ATE in the target population $\tilde{\tau}^\ast := \E_{P^\ast}\left\{Y(1) - Y(0)\right\} = \E_{P^\ast}\left[E_{P^\ast}\left\{Y(1) - Y(0) \mid X\right\}\right]$, a standard estimand in the transportability and generalizability literature. However, we do not use Assumption \ref{as:general}, nor do we make $\tilde{\tau}^\ast$ our estimand, because doing so would in effect assume away the phenomenon we seek to isolate. That is, rather than assuming that observed unit-level covariate heterogeneity explains all treatment effect variation across populations, we seek to develop a realistic account of the extent to which it explains treatment effect variation relative to other possible factors.

\subsection{Related Work}
\label{sec:related}
This paper builds on a growing literature on the transportability and generalizability of treatment effects. For the most part, the setup and assumptions established in Section \ref{sec:assumptions} are common in this literature. See, for example, \citet{hotz2005}, \citet{allcott2015}, \citet{crepon2018}, \citet{dahabreh2018}, and \citet{egamielements2021}.

However, two features of our work are worth emphasizing. First, we consider identification and estimation of site ATEs transported to an arbitrary target population. This target population can be finite or infinite. And it can but need not contain the experimental population. By contrast, previous work mostly focuses on transportation to an observed, finite target population of units and specifies whether the target population contains the experimental population or not \citep{hotz2005, dahabreh2018, dahabreh2018c}. Second, as we note above, we seek to identify and estimate $\tau^\ast$ rather than $\tilde{\tau}^\ast$. So we do not require Assumption \ref{as:general}, which is often quite strong and difficult to verify in practice.

Despite these two differences, we can naturally adapt standard estimators for transported ATEs to estimate $\tau^\ast$ in our present setup. These estimators, which can be obtained by plug-in estimation of the terms that identify $\tau^\ast$ in (\ref{eq:identify}), generally mirror standard outcome-modeling and IPW estimators from the observational causal inference literature; see  \citet{ackerman2019implementing} and \citet{egamielements2021} for recent discussion.
For example, if the conditional mean outcome model $m_{z}(x) := \E_{P}\left\{Y(z) \mid X = x\right\}$ is correctly specified for $z \in \{0, 1\}$, then the outcome-modeling estimator
\begin{equation}
    \hat{\tau}^{\ast}_{\text{om}} := \E_{P^\ast}\left\{\hat{m}_{1}(X) - \hat{m}_{0}(X)\right\}
\end{equation}
is consistent for $\tau^\ast$ \citep[see, for example,][]{kern2016assessing} when the distribution $P^\ast$ is known.
Additionally, the IPW estimator 
\begin{equation}
\label{eq:ipw}
    \hat{\tau}^\ast_{\text{ipw}} := \frac{1}{n}\sum_{i = 1}^{n} \widehat{\frac{dP^\ast}{dP}}(X_i)\left\{\frac{Z_i}{\pi} Y_i - \frac{1 - Z_i}{1-\pi}Y_i\right\}
\end{equation}
is consistent for $\tau^\ast$ if 
$\widehat{\frac{dP^\ast}{dP}}(x)$ is correctly specified.
When both the experimental and target populations are finite, we can estimate $\widehat{\frac{dP^\ast}{dP}}(x)$ as follows. Let $E \in \{0, 1\}$ indicate inclusion in the experimental population and $T \in \{0, 1\}$ indicate inclusion in the target population. By Bayes' rule, the change in measure is
\begin{equation}
\label{eq:bayes}
\frac{dP^\ast}{dP}(x) = \frac{p^\ast(x)}{p(x)} = \frac{\Pr(T = 1 \mid X = x)}{\Pr(E = 1 \mid X = x)}\frac{\Pr(E = 1)}{\Pr(T = 1)}.
\end{equation}
We can then separately estimate the conditional probabilities in \eqref{eq:bayes}. See \citet{westreich2017} and \citet{dahabreh2018c} for analogous discussion in a similar setting.
Finally, the doubly robust estimator
\begin{equation}
\begin{aligned}
    \hat{\tau}^{\ast}_{\text{dr}} := &\left[\frac{1}{n}\sum_{i = 1}^{n}\widehat{\frac{dP^\ast}{dP}}(X_i)\frac{Z_i}{\pi}\left\{Y_i - \hat{m}_{1}(X_i)\right\} + \E_{P^\ast}\left\{\hat{m}_{1}(X)\right\}\right]\\
    &- \left[\frac{1}{n}\sum_{i = 1}^{n}\widehat{\frac{dP^\ast}{dP}}(X_i)\frac{1 - Z_i}{1 - \pi}\left\{Y_i - \hat{m}_{0}(X_i)\right\} + \E_{P^\ast}\left\{\hat{m}_{0}(X)\right\}\right]
\end{aligned}
\end{equation}
is consistent for $\tau^\ast$ if either the outcome model or the change of measure model is correctly specified.
Targeted maximum likelihood estimators for transported site ATEs \citep{rudolph2017} can similarly be adapted to our setup, but we do not discuss them at length.

IPW and outcome-modeling estimators, however, can perform poorly in many settings. IPW can produce extreme estimates when there is limited overlap between the experimental and target populations or when there are many covariates \citep{kang2007demystifying}. 
And outcome modeling can also perform poorly in moderate to high dimensions, when the performance of more flexible outcome models can start to degrade \citep[see][]{kern2016assessing}.

Our proposed weighting approach draws on the recent literature on approximate balancing weights in observational causal inference. See, for example, \citet{Zubizarreta2015, Athey2018, hirshberg2020}. \citet{benmichael2021review} give a recent review. These methods find weights that minimize a measure of covariate imbalance between two groups of units, typically between treated and control units in observational studies. A special case of this approach called entropy balancing \citep{hainmueller2012} arises when the weights can achieve \emph{exact balance} in the covariates. In a set of papers closely aligned with ours, \citet{josey2020exp} and \citet{josey2020} propose adapting entropy balancing to transport treatment effects. These papers offer an important conceptual advance relative to transportation methods that rely on traditional IPW. However, weights that achieve exact balance are often infeasible in practice, even with a moderate number of covariates. Thus, our proposed approach extends their basic framework to allow for \emph{approximate balance}, following a suggestion in the conclusion of \citet{josey2020exp}. Finally, for a related setting, \citet{crepon2018} propose using outcome modeling for dimension reduction and then entropy balancing for transportation. It is straightforward to adapt their approach to use approximate balancing weights, as proposed here, rather than entropy balancing.

\section{Weighting Estimators for Transported Site Average Treatment Effects}
\label{sec:estimation}

We propose estimating $\tau^\ast$ via a linear weighting estimator $\hat{\tau}^\ast$ with weights $\hat{\gamma} \in \R^n$:
\begin{equation}
    \hat{\tau}^\ast := \frac{1}{n}\sum_{i = 1}^{n}\hat{\gamma}_i \frac{Z_i}{\pi} Y_i - \frac{1}{n}\sum_{i = 1}^{n}\hat{\gamma}_i\frac{1 - Z_i}{1-\pi}Y_i = \frac{1}{n}\sum_{i=1}^n \hat{\gamma}_i \frac{ Z_i - \pi}{\pi(1-\pi)} Y_i.
\end{equation}
Recall that we assume for clarity of exposition that the propensity score is constant: $e(X_i) = \pi$.
This is satisfied by design in many simple experimental setups. The methods we propose in this section can be extended by analogy to non-constant propensity scores.

Intuitively, we want to choose weights that optimize the performance of $\hat{\tau}^\ast$ as an estimator for $\tau^\ast$ by some metric. In this paper, we specifically consider choosing weights to minimize the estimation error given by $\hat{\tau}^{\ast} - \tau^\ast$, adapting arguments and estimators from \citet{Athey2018} and \citet{hirshberg2020} designed for observational studies to estimate transported effects. We discuss in Section \ref{sec:estimation_error} one way of decomposing this estimation error, then show in Section \ref{sec:min_max} how we can use convex optimization to choose weights that minimize in a controlled way the constituent parts of the decomposed estimation error.

\subsection{Estimation Error Decomposition}
\label{sec:estimation_error}
First, we decompose the estimation error $\hat{\tau}^\ast - \tau^\ast$ into three terms: (1) covariate imbalance between treated and control units in the experimental population, (2) covariate imbalance between treated units in the experimental population and the target population, and (3) error due to noise. Doing so yields

\begin{equation}
    \label{eq:est_error}
    \begin{aligned}
     \hat{\tau}^\ast - \tau^\ast & = \underbrace{\frac{1}{n} \sum_{i=1}^n \hat{\gamma}_i \frac{Z_i - \pi}{\pi (1-\pi)} m_0(X_i)}_{\text{imbalance in } m_0(\cdot)}  + \underbrace{\frac{1}{n\pi}\sum_{i=1}^n \hat{\gamma}_iZ_i\tau(X_i) - \E_{P^\ast}\{\tau(X)\}}_{\text{imbalance in } \tau(\cdot)} + \underbrace{\frac{1}{n}\sum_{i=1}^n \hat{\gamma}_i \frac{Z_i - \pi}{\pi (1-\pi)}\varepsilon_i}_{\text{noise}},
    \end{aligned}
\end{equation}
where $\varepsilon_i := Y_i - m_0(X_i) - Z_i\tau(X_i)$.
The first term in \eqref{eq:est_error} is the imbalance in the expected control potential outcome $m_0(\cdot)$ \emph{within the study}. It is largely controlled by design via treatment randomization; if the weights $\hat{\gamma}$ are simply uniform, then this term will be zero in expectation by Assumption \ref{as:random}.
However, any particular experiment will have some chance imbalance. In Section \ref{sec:min_max}, we propose choosing weights $\hat{\gamma}$ that adjust for this in a manner similar to regression adjustment or post-stratification \citep{Lin2013}. 
The second term in \eqref{eq:est_error} measures the discrepancy between the treatment effect in the treated experimental population and the treatment effect in the target population $P^\ast$. We are primarily interested in controlling this term. The final term in \eqref{eq:est_error} arises from noise in the outcomes and is related to the variance of the estimator.

If the prognostic score $m_0(\cdot)$ is in model class $\calM$ and the CATE $\tau(x)$ is in model class $\calT$, then we can bound the estimation error by 
\begin{equation}
    \label{eq:error_bound}
    \begin{aligned}
     \left|\hat{\tau}^\ast - \tau^\ast\right| & \leq \sup_{m \in \calM} \left|\frac{1}{n} \sum_{i=1}^n \hat{\gamma}_i \frac{Z_i - \pi}{\pi (1-\pi)} m(X_i)\right|  + \sup_{\tau \in \calT} \left|\frac{1}{n\pi}\sum_{i=1}^n \hat{\gamma}_iZ_i\tau(X_i) - \E_{P^\ast}\{\tau(X)\}\right| + \left|\frac{1}{n}\sum_{i=1}^n \hat{\gamma}_i \frac{Z_i - \pi}{\pi (1-\pi)}\varepsilon_i\right|.
    \end{aligned}
\end{equation}

\subsection{Minimizing the Worst-case Estimation Error}
\label{sec:min_max}
To control the estimation error \eqref{eq:est_error} given model classes for the prognostic score $\calM$ and the CATE $\calT$, we use convex optimization to find weights that solve
\begin{equation}
\label{eq:primal}
\begin{aligned}
     \min_\gamma \;\; & \sup_{m_0 \in \calM_0} \left|\frac{1}{n}\sum_{i=1}^n\gamma_i\frac{Z_i - \pi}{\pi (1-\pi)}m_0(X_i)\right|^2\\
     &  + \sup_{\tau \in \calT} \left|\frac{1}{n\pi}\sum_{i=1}^n\gamma_i Z_i \tau(X_i) - \E_{P^\ast}\{\tau(X)\}\right|^2 +
     \lambda\sum_{i=1}^n\gamma_i^2\left(\frac{Z_i}{\pi} + \frac{1-Z_i}{1-\pi}\right)\\
     \text{subject to   } & \sum_{i=1}^n Z_i \gamma_i = n_1, \;\;\;\; \sum_{i=1}^n (1-Z_i) \gamma_i = n_0, \;\;\;\; \gamma_i \geq 0.
\end{aligned}
\end{equation}
The objective in optimization problem \eqref{eq:primal} directly targets the upper bound on the estimation error \eqref{eq:error_bound}. First, the optimization problem finds weights that optimize external validity by weighting treated units so that the weighted CATE functions resemble the target distribution's treatment effect $\tau^\ast$ for the worst-case CATE function. At the same time, the optimization problem maintains internal validity by weighting the control units so that they are comparable to the weighted distribution of treated units with respect to the prognostic score $m_0(\cdot)$. The final term in the objective penalizes the weights for non-uniformity via an $L^2$ regularization term; this is a proxy for the variance due to noise in \eqref{eq:error_bound}. We include a regularization hyper-parameter $\lambda$ that controls this tradeoff between better balance (lower bias) and more uniform weights (lower variance). We describe the choice of $\lambda$ in practice in Section \ref{sec:app}.

The optimization problem \eqref{eq:primal} includes three constraints for stability of the estimator. The first two constrain the weights on the treated units to sum to the number of treated units $n_1$ and constrain the weights on the control units to sum to the number of control units $n_0$. This ensures that the estimator is robust to constant shifts in the outcome \citep{benmichael2021review}.
The third constraint restricts weights to be non-negative. This prohibits extrapolation from the support of the experimental sample when estimating $\tau^\ast$ \citep[see][]{king2006dangers, Zubizarreta2015, benmichael2020_augsynth}. Taken together, these constraints also ensure that the estimator is sample-bounded.

Denoting $\hat{\mu}_1^\ast = \frac{1}{n_1}\sum_{i=1}^n Z_i \hat{\gamma}_i Y_i$ and $\hat{\mu}_0^\ast = \frac{1}{n_0} \sum_{i=1}^n (1-Z_i) \hat{\gamma}_i Y_i$ as the weighted average of treated and control units, we estimate standard errors via the heteroskedasticity-robust sandwich estimator
\begin{equation}
    \label{eq:standard_error}
    \hat{V} = \frac{1}{n_1 - 1}\sum_{i=1}^n Z_i \hat{\gamma}_i^2(Y_i - \hat{\mu}_1^\ast)^2 + \frac{1}{n_0 - 1}\sum_{i=1}^n (1-Z_i) \hat{\gamma}_i^2(Y_i - \hat{\mu}_0^\ast)^2.
\end{equation}
In defining $\hat{V}$, we use the constraints that the weights over the treated units sum to $n_1$ and the weights over the control units sum to $n_0$. Note that, with uniform weights, \eqref{eq:standard_error} is the usual variance estimator for the treatment effect in a randomized trial. See \citet{hirshberg2020} for discussion of asymptotic normality and efficiency of estimators of this form.

\subsection{Implementation}

We conclude by discussing how we can implement the optimization problem \eqref{eq:primal} for two pairs of model classes when the target distribution $P^\ast$ is the empirical distribution over $m$ units with covariates $\{\widetilde{X}_1,\ldots,\widetilde{X}_m\}$.
In the first, both $m_0(\cdot)$ and $\tau(\cdot)$ are linear in transformations of the observed covariates: $\calM = \{\beta_0 \cdot \phi_0(x)\mid \|\beta_0\| \leq C_0\}$ and $\calT = \{\beta_\tau \cdot \phi_{\tau}(x) \mid \|\beta_\tau\| \leq C_\tau\}$.
 In this formulation, we allow the models to use different transformations of the covariates so that, \textit{e.g.}, $m_0(\cdot)$ has a more flexible basis expansion while $\tau(x)$ is restricted to a simpler CATE function. \citep[See][for related discussion.]{Kunzel2019, Hahn2020}
By the Cauchy-Schwarz inequality, the special case of optimization problem \eqref{eq:primal} when we use the $L^2$ norm $\|\cdot\|_2$ in the objective function is
\begin{equation}
\label{eq:primal_linear}
\begin{aligned}
     \min_\gamma \;\; & \left\|\frac{1}{n\pi}\sum_{i=1}^n\gamma_i Z_i \phi_\tau(X_i) - \frac{1}{m}\sum_{\ell=1}^m\phi_\tau(\widetilde{X}_\ell)\right\|_2^2 + \left\|\frac{1}{n}\sum_{i=1}^n\gamma_i\frac{Z_i - \pi}{\pi (1-\pi)}\phi_0(X_i)\right\|_2^2 +
     \lambda\sum_{i=1}^n\gamma_i^2\left(\frac{Z_i}{\pi} + \frac{1-Z_i}{1-\pi}\right)\\
     \text{subject to   } & \sum_{i=1}^n Z_i \gamma_i = n_1, \;\;\;\; \sum_{i=1}^n (1-Z_i) \gamma_i = n_0, \;\;\;\; \gamma_i \geq 0.
\end{aligned}
\end{equation}
This optimization problem is a quadratic program (QP), so it can be efficiently solved even with many units and high-dimensional $\phi_0(X_i)$ and $\phi_\tau(X_i)$---by, for example, the alternating direction method of multipliers \citep{Boyd2010}. Note that reweighting to the target distribution requires only the \emph{average} of $\phi_\tau(\widetilde{X}_\ell)$ over the target distribution and we can solve \eqref{eq:primal_linear} without access to the full underlying distribution $\{\widetilde{X}_1,\ldots,\widetilde{X}_m\}$. This reduced data requirement can be useful in practice.

Second, we consider a generalization to linearity with infinite-dimensional transformations $\phi_0(\cdot)$ and $\phi_\tau(\cdot)$. Let $\calH_{k_0}$ and $\calH_{k_{\tau}}$ be reproducing kernel Hilbert spaces with associated kernels $k_0: \R^d \times \R^d \to \R$ and $k_\tau:\R^d \times \R^d \to \R$. Then we can consider the model class for the prognostic score to be $\calM = \{m_0(\cdot) \in \calH_{k_0} \mid \|m_0\|_{\calH_{k_0}} \leq 1\}$ and for the CATE to be $\calT = \{\tau(\cdot) \in \calH_{k_\tau} \mid \|\tau\|_{\calH_{k_\tau}} \leq 1\}$. By the reproducing property, with these model classes we can write the prognostic score and the CATE as linear functions $m_0(x) = \<m_0(\cdot), k_0(x, \cdot)\>$ and $\tau(x) = \<\tau(\cdot), k_{\tau}(x, \cdot)\>$, where the representations $\phi_0(x) = k_0(x, \cdot)$ and $\phi_\tau(x) = k_\tau(x, \cdot)$ are defined by the kernel functions $k_0$ and $k_\tau$, respectively. As above, by H\"{o}lder's inequality the suprema in \eqref{eq:error_bound} are the imbalances in the \emph{infinite}-dimensional transformations of the covariates, $\phi_0(X_i)$ and $\phi_\tau(X_i)$. Using the ``kernel trick''---$\<\phi_0(x), \phi_0(y)\> = k_0(x,y)$ and $\<\phi_\tau(x), \phi_\tau(y)\> = k_\tau(x, y)$---we can write the balancing weights optimization problem as
\begin{equation}
\label{eq:primal_kernel}
\begin{aligned}
     \min_\gamma \;\; & \frac{1}{n^2\pi^2}\sum_{i,j=1}^n \gamma_i\gamma_j Z_i Z_j k_\tau(X_i, X_j)  - 2 \frac{1}{nm\pi}\sum_{i=1}^n\sum_{\ell=1}^m \gamma_i Z_i k_\tau(X_i, \widetilde{X}_\ell)\\
     & +\frac{1}{n^2}\sum_{i=1}^n\sum_{j=1}^n \gamma_i\gamma_j\frac{Z_i - \pi}{\pi(1-\pi)} \frac{Z_j - \pi}{\pi(1-\pi)} k_0(X_i, X_j) +
     \lambda\sum_{i=1}^n\gamma_i^2\left(\frac{Z_i}{\pi} + \frac{1-Z_i}{1-\pi}\right)\\
     \text{subject to   } & \sum_{i=1}^n Z_i \gamma_i = n_1, \;\;\;\; \sum_{i=1}^n (1-Z_i) \gamma_i = n_0, \;\;\;\; \gamma_i \geq 0.
\end{aligned}
\end{equation}
As above, this optimization problem is a QP. 
To compute the imbalance terms, we only need to compute the kernel evaluations $k_0(X_i, X_j)$ and $k_\tau(X_i, X_j)$ in the experimental population, as well as the expected kernel evaluation in the target distribution $\E_{P^\ast}\{k(X_i, X)\}$ for each treated experimental unit $X_i$.
However, unlike in \eqref{eq:primal_linear}, the full set of unit-level covariates in the target population $\{\widetilde{X}_1,\ldots,\widetilde{X}_m\}$ are needed in \eqref{eq:primal_kernel} to compute the kernel evaluations between the treated units and the target population, $k_\tau(X_i, \widetilde{X}_\ell)$. Solutions to both of these optimization problems can be found using the \texttt{osqp} QP solver \citep{osqp}.

\section{Extension to Multisite RCTs}
\label{sec:multi}

Consider a multisite RCT with $n$ total units across $J$ sites, in which the $n_j$ units in site $j$ are separately randomized with probability of treatment $\pi_j$.
To extend our framework to this setting, we simply repeat the analysis in Sections \ref{sec:setup} and \ref{sec:estimation} separately for each of the $J$ sites.
Our estimand for the $j^{\text{th}}$ site is the ATE in the site if its population had the same distribution of observed unit-level covariates as the target population:
\begin{equation}
\label{eq:multisite_estimand}
    \tau^{\ast}_{j} := \E_{P^\ast}\left[\E_{P}\left\{Y(1) - Y(0) \mid X, S = j\right\}\right]
    = \E_{P^\ast}\left\{m_{1j}(X) - m_{0j}(X)\right\}
    = \int \tau_j(x)\ dP^\ast(x),
\end{equation}
where $S \in \{1, \ldots, J\}$ indicates the unit's site membership, $m_{zj}(x) := \E_{P}\left\{Y(z) \mid X = x, S = j\right\}$ is the conditional mean potential outcome function in the $j^{\text{th}}$ site, and $\tau_j(x) = m_{1j}(x) - m_{0j}(x)$ is the CATE in the $j^{\text{th}}$ site.
If Assumptions \ref{as:random}--\ref{as:cont} hold within each site, then $\tau^{\ast}_{j}$ is identifiable as in (\ref{eq:identify}) for each $j \in \{1, \ldots, J\}$.

Our weighting approach to estimating $\tau^\ast$ in the single-site setting can be directly applied to estimate $\tau^{\ast}_j$ for any $j \in \{1, \ldots, J\}$: We simply treat the units in the $j^{\text{th}}$ site as the experimental population and ignore experimental units in other sites.
We thus refer to our weighting estimator in this multisite context as $\hat{\tau}^{\ast}_j$.
By sequentially solving (\ref{eq:primal}) for each $j \in \{1, \ldots, J\}$, we can obtain the weights $\hat{\gamma}$ that define our weighting estimators $\hat{\tau}^{\ast}_1, \ldots, \hat{\tau}^{\ast}_J$.
The standard outcome-modeling, IPW, and doubly robust estimators from Section \ref{sec:related} can similarly be extended to this multisite setting by restricting attention only to the units from the site of interest $j$.

We can then use this approach to decompose treatment effect variation across the site effects.
In particular, researchers often want to understand how much treatment effect variation is attributable to differences in observed unit-level covariate distributions across sites, as opposed to differences in unobserved unit-level covariate distributions or differences in site-level features. 
Direct comparisons of the site ATEs $\tau_1, \ldots, \tau_J$ typically cannot disaggregate these sources of variation, since the site ATEs incorporate all sources of variation simultaneously.
But having access to $\tau^{\ast}_j$ for different sites $j$ and target distributions $P^\ast$, in addition to the untransported $\tau_j$, can help answer questions like:
\begin{enumerate}
    \item What would the distribution of site ATEs be if all sites had the same distribution of observed unit-level covariates?
    \item What is the relationship between the site ATEs and site-level covariates, net of differences in unit-level covariate distributions between sites?
    \item How much variation in site ATEs can be explained by the observed unit-level covariates?
\end{enumerate}
Data from multisite RCTs are rich enough to answer these questions because treatment is randomized in all sites, so the CATE function $\tau_j(x)$ of each site is identifiable.
The first question can be answered by examining the distribution of $\tau^\ast_1, \ldots, \tau^\ast_J$ for a fixed target distribution $P^\ast$---\textit{i.e.}, by integrating each site's CATE function over the same target covariate distribution.
The second question can be answered by a descriptive analysis of the relationship between site-level covariates and estimates of $\tau^\ast_1, \ldots, \tau^\ast_J$ for a fixed $P^\ast$. 

To answer the third question, we compare the variability of $\tau_1, \ldots, \tau_J$ to that of $\tau^{\ast}_1, \ldots, \tau^{\ast}_J$. We can think of the change in variability after transportation as an $R^2$-type measure that approximates the amount of variability in site ATEs attributable to differences in observed unit-level covariate distributions. Heuristically, if $\theta$ and $\theta^\ast$ denote the finite-sample variances of $\tau_1, \ldots, \tau_J$ and $\tau^{\ast}_1, \ldots, \tau^{\ast}_j$, then the pseudo-$R^2$ measure
\begin{equation}
\label{eq:r2}
    R^2 := 1 - \left(\frac{\theta^\ast}{\theta}\right)^2
\end{equation}
approximates the amount of variability attributable to observed unit-level covariates.
This is only a pseudo-$R^2$ measure since variability might increase after transportation ($\theta^\ast > \theta$) if differences in observed unit-level covariate distributions were masking variability due to differences in site-level features.

To estimate $\theta$ and $\theta^\ast$, we use the ``$Q$-statistic'' approach from the meta-analysis literature \citep[see, \textit{e.g.},][]{hedges2001}. \citet{keele2020} apply it in a similar setting. An advantage of the approach is that it isolates the variability of the true site ATEs from the variability of their estimates. This is especially important in our context since weighting tends to reduce effective sample size and thus increase estimation variability. Given a hypothesized amount of variability in site ATEs $\theta$, the $Q$-statistic is
\begin{equation}
\label{eq:q}
    Q(\theta) := \sum_{j = 1}^{J}\frac{\left(\hat{\tau}_j - \overline{\tau}\right)^2}{\widehat{\text{se}}^2_j + \theta},
\end{equation}
where $\overline{\tau}$ is the estimated overall ATE and $\widehat{\text{se}}^2_j$ is the estimated standard error of the estimated ATE in the $j^{\text{th}}$ site $\hat{\tau}_j$.\footnote{For intuition on the terms in the denominator of \eqref{eq:q}, note that $\hat{\tau}_{j} - \overline{\tau} = (\hat{\tau}_{j} - \tau_j) - (\overline{\tau} - \tau_j)$. Roughly speaking, $\widehat{\text{se}}_j^2$ reflects the variation due to estimation error, $(\hat{\tau}_{j} - \tau_j)$. And $\theta$ reflects the variation due to deviations of $\tau_j$ from $\overline{\tau}$, $(\overline{\tau} - \tau_j)$.} 

Under the null hypotheses that $\theta = \theta_0$, $Q(\theta_0)$ approximates a $\chi^2_{J - 1}$ distribution. So we can estimate $\theta$ using a Hodges-Lehman point estimate (corresponding to the $\theta$ with the largest $p$-value, realized at $Q(\theta) = J-1$), and generate a confidence interval via test inversion. By applying this procedure for both the untransported site ATEs $\tau_1, \ldots, \tau_J$ and the transported site ATEs $\tau^{\ast}_{1}, \ldots, \tau^{\ast}_J$, we can estimate the variability in site ATEs before and after transportation. We then plug these estimates into \eqref{eq:r2} to get our pseudo-$R^2$ measure.

We demonstrate how the above analyses can be conducted in our study of welfare-to-work experiments in Section \ref{sec:app}.

\section{Simulation Study}
\label{sec:sim}
In this section, we examine the behavior of our proposed estimators and compare them to outcome-modeling, IPW, and doubly robust estimators via simulation. We base our simulation on data from the welfare-to-work experiments analyzed by \citet{bloom2001} and \citet{bloom2003}, introduced in Section \ref{sec:intro}. Our outcome of interest is log earnings; we add 100 before logging to avoid undefined values.

We generate data for our simulation using this dataset as follows. First, we define the CATE $\tau_j(x)$ in the $j^{\text{th}}$ site to be a sparse linear function of unit-level covariates $X_i$ plus an intercept term specific to the multisite experiment to which the $j^{\text{th}}$ site belongs. For most covariate values observed in the data, $\tau_j(x)$ was less than 1 in magnitude. Its most extreme value was $-4.5$. In each simulation repetition, we generate a bootstrap sample of each of the 59 sites, define the potential outcomes under control $Y_i(0)$ to be the observed outcomes of the bootstrapped units plus a noise term $\epsilon_i \simiid \calN\left(0, 0.5^2\right)$, and generate potential outcomes under treatment $Y_i(1)$ by adding $\tau_j(X_i)$ and another independent noise term $\delta_i \simiid \calN\left(0, 0.5^2\right)$ to $Y_i(0)$. We then randomize treatment assignment, keeping the proportion of treated units in each site the same as in the original dataset, and set the observed outcome to be the corresponding potential outcome. Thus generated, the observed and potential outcomes had standard deviations of about 2.3.

In each repetition, our estimands are the site ATEs transported to the bootstrapped population of 46,977 units in the NEWWS experiment. In other words, we seek to estimate $\tau^{\ast}_1, \ldots, \tau^{\ast}_{59}$ defined in (\ref{eq:multisite_estimand}), where $P^{\ast}$ is the bootstrapped population of units from NEWWS. Note that, for any two sites $k$ and $\ell$, $\tau^{\ast}_k$ and $\tau^{\ast}_\ell$ differ only by their experiment-specific intercepts since the unit-level covariates define $\tau_j(x)$ in the same way for all $j \in \{1, \ldots, 59\}$ by design. To be concrete, these estimands ranged between about $-1.25$ and $0.3$.

We estimate $\tau^{\ast}_1, \ldots, \tau^{\ast}_{59}$ using two versions of our weighting estimator $\hat{\tau}_{j}^{\ast}$. The first, which we refer to as ``linear,'' optimizes the weights over the class of linear functions of $X_i$ that could define $m_{0j}(x)$ and $\tau_j(x)$, as in (\ref{eq:primal_linear}). The second version, which we refer to as ``RBF,'' optimizes the weights to solve \eqref{eq:primal_kernel} where the kernel function for the control potential outcomes, $k_0(\cdot, \cdot)$, is the radial basis function (RBF) kernel, and the kernel for the CATE, $k_\tau(\cdot, \cdot)$ is linear. The linear and RBF versions of $\hat{\tau}_{j}^{\ast}$ both require only the covariate means, not the full covariate distribution, of the target population. And they both correctly specify $\tau_j(x)$. But the RBF version of $\hat{\tau}_{j}^{\ast}$ allows a more flexible model for $m_{0j}(x)$ than the linear version does. Since only one covariate is continuous, however, it is not obvious that this greater flexibility will appreciably advantage the RBF version.

We also estimate $\tau^{\ast}_1, \ldots, \tau^{\ast}_{59}$ using the 
outcome-modeling, 
IPW, and doubly robust estimators described in Section \ref{sec:related}. For these models, we use logistic regression to estimate $\Pr(S = j \mid X)$ and linear regression to estimate $m_{zj}(x)$. Finally, as a benchmark for comparison, we also use the naive, unadjusted difference-in-sample-means estimate of each site ATE
\[
\frac{1}{\pi_jn_j}\sum_{i:S_i = j}Z_iY_i(1) - \frac{1}{(1 - \pi_j)n_j}\sum_{i:S_i = j}(1 - Z_i)Y_i(0), \hspace{20pt} j = 1, \ldots, 59
\]
to estimate $\tau^{\ast}_1, \ldots, \tau^{\ast}_{59}$.

\begin{figure}
    \centering
    \includegraphics[width = 0.99\linewidth]{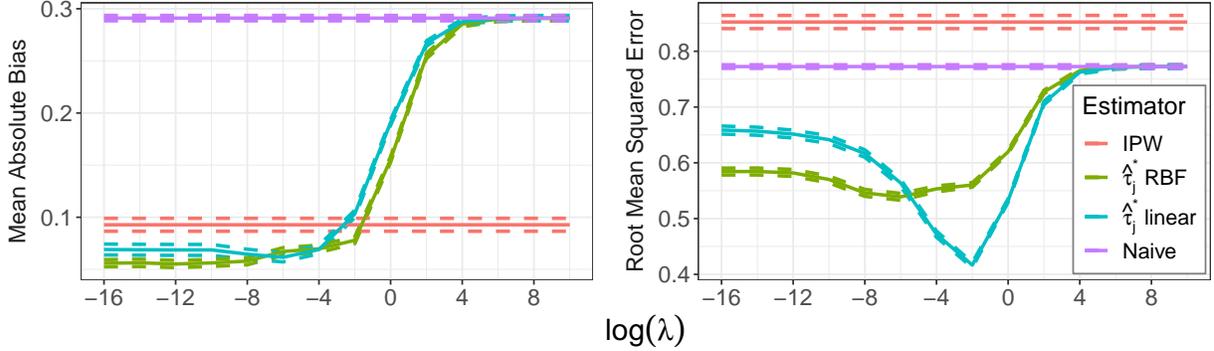}
    \caption{Mean absolute bias and root mean squared error of each estimator over 120 repetitions of the welfare-to-work simulation. Observed outcomes are used as potential outcomes under control, and the CATE is defined as a sparse linear function of unit-level covariates plus an experiment-specific intercept term. Results for the outcome-modeling and doubly robust estimators are omitted because they are an order of magnitude larger. For comparison, the estimands $\tau^\ast_1, \ldots, \tau^\ast_{59}$ are no greater in magnitude than 1.3, and the CATEs are no greater in magnitude than 4.5.}
    \label{fig:mdrc_sim_bias_rmse_C_site}
\end{figure}

We evaluate the estimators' performance by computing in each repetition their root mean squared errors (RMSEs) and biases over the 59 sites. Figure \ref{fig:mdrc_sim_bias_rmse_C_site} shows the RMSE and mean absolute bias of each estimator, averaged over the simulation repetitions. To help interpret the magnitudes of these graphs, we note that the estimands $\tau^\ast_1, \ldots, \tau^\ast_{59}$ are no greater in magnitude than 1.3, and the CATEs are no greater in magnitude than 4.5. The RMSEs and absolute biases of the outcome-modeling and doubly robust estimators are not shown because they are an order of magnitude larger than those of the estimators shown. Upon closer inspection, we determined that this is due to limited covariate overlap between certain sites and the target.

We note four salient features of Figure \ref{fig:mdrc_sim_bias_rmse_C_site}. First, under a range of regularization levels, our weighting estimators have lower RMSEs and mean absolute biases than the other estimators. Second, our weighting estimators are least biased when almost completely unregularized. Third, the RMSEs of our weighting estimators are lowest at some medium level of regularization. Finally, our weighting estimators reduce to the naive estimator when heavily regularized. These trends illustrate the bias-variance tradeoff of regularization.

\section{Empirical Application}
\label{sec:app}

\begin{figure}
    \centering
    \includegraphics[width = 0.99\linewidth]{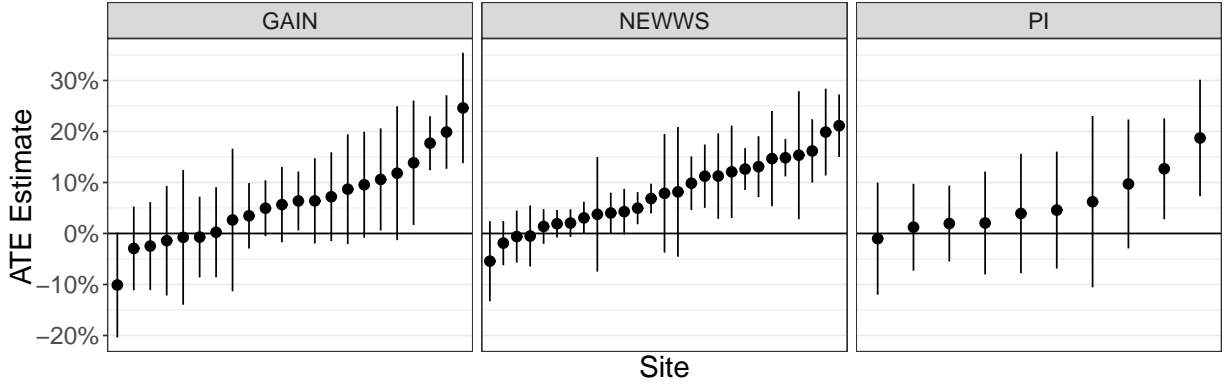}
    \caption{Point estimate and 95\% confidence interval for the untransported ATE in each site.}
    \label{fig:raw_ates}
\end{figure}

In this section, we apply our method to characterize treatment effect variation in the welfare-to-work experimental data. Our outcome of interest in this analysis is a binary indicator of employment two years after treatment assignment, measured by nonzero earnings. Figure \ref{fig:raw_ates} plots the point estimate and 95\% confidence interval for the ATE in each site. 
In many sites, assignment to welfare-to-work programs increased the probability of employment two years later.
However, interpreting these site-specific estimates collectively is difficult because there are several different sources of treatment effect variation across sites. For example, a welfare-to-work program might be more effective in urban sites. But it also might be more effective for non-Hispanic white people, who are more common in rural and suburban sites than in urban ones. A principled framework for cross-site effect comparisons is needed to disentangle these sources of variation.

We adopt the framework from Section \ref{sec:multi}. Under this framework, we first estimate what the ATE in each site would be if all sites had the same population of units. We can then attribute the remaining treatment effect variation across sites to differences in site-level or unobserved unit-level covariates. In this case, we transport the ATE estimates to the overall population of 69,399 units. 
We do so by solving the optimization problem (\ref{eq:primal_linear}), where $\phi_{\tau}(x)$ and $\phi_{0}(x)$ each map the 23 observed covariates to themselves and a few of their interactions, such as the interactions between race and baseline income bracket. We standardize the covariates and their interactions to have unit variance.

As discussed in Section \ref{sec:estimation}, we must choose the regularization parameter $\lambda$ in (\ref{eq:primal_linear}). To do so, we first solve (\ref{eq:primal_linear}) for a range of values of $\lambda$, then compare the resulting reductions in covariate imbalance and effective sample size. Given weights $\hat{\gamma}$, we measure covariate imbalance between the treated and control groups by
\[
\frac{1}{59}\sum_{j = 1}^{59}\left\|\frac{1}{n_j}\sum_{i:S_i = j}\hat{\gamma}_{i}\frac{Z_i - \pi_j}{\pi_j(1 - \pi_j)}\phi_{0}(X_i)\right\|_2.
\]
We measure covariate imbalance between the treated group and the target population by
\[
\frac{1}{59}\sum_{j = 1}^{59}\left\|\frac{1}{n_j\pi_j}\sum_{i:S_i = j}\hat{\gamma}_{i}Z_{i}\phi_{\tau}(X_i) - \mathbb{E}_{P^\ast}\left\{\phi_{\tau}(X)\right\}\right\|_2.
\]
To measure effective sample size, we compute Kish's effective sample size in each site \citep{kish1965}, then average those values over the sites.

\begin{figure}
    \centering
    \includegraphics[width = 0.99\linewidth]{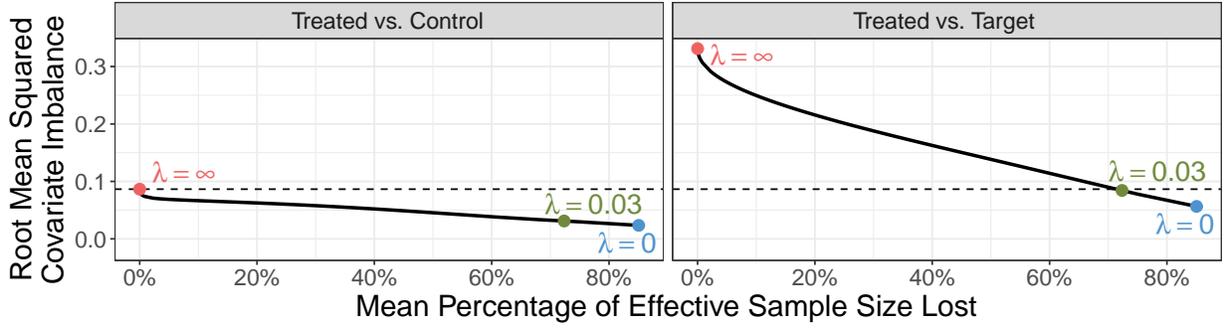}
    \caption{Reductions in covariate imbalance and effective sample size due to weighting with different levels of regularization. The dashed line marks the level of covariate imbalance observed between the treated and control groups without weighting. This imbalance is due to chance randomization alone. Regularization manages the tradeoff between covariate imbalance (bias) and effective sample size (variance).}
    \label{fig:imbalance_sample_size}
\end{figure}

Figure \ref{fig:imbalance_sample_size} shows the tradeoff between imbalance reduction and effective sample size reduction. Since treatment was randomized in each site, the treated and control groups were fairly balanced even before weighting. So we focus on covariate imbalance between the treated group and the target distribution. Setting $\lambda = 0.03$ reduces covariate imbalance by 80\%. The remaining imbalance is less than the imbalance between the unweighted treated and control groups, which, recall, is due to chance randomization alone. This imbalance reduction comes at the cost of an approximately 70\% reduction in effective sample size.

\begin{figure}
    \centering
    \includegraphics[width = 0.99\linewidth]{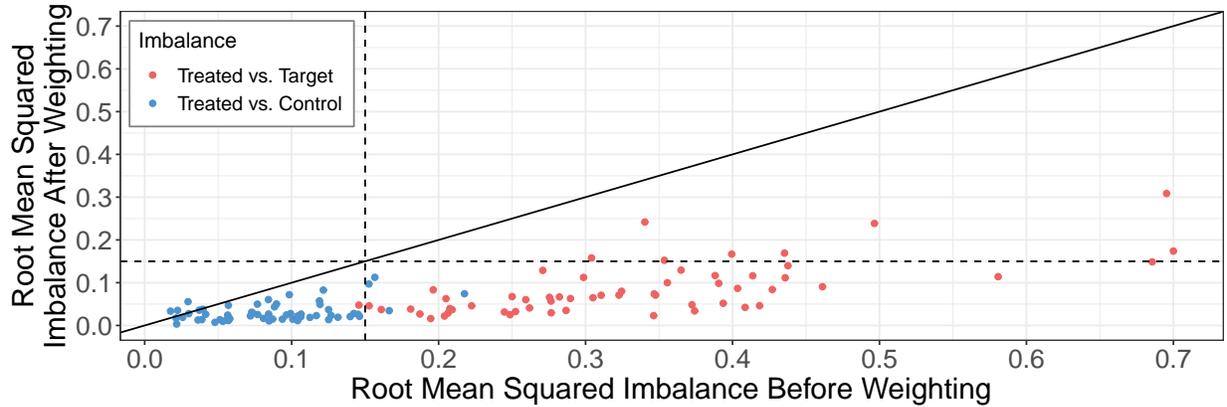}
    \caption{Root mean squared covariate imbalance between the treated group and the control group (blue) and between the treated group and the target population (red) before and after weighting. Note that the covariates were standardized to have unit variance. In most sites, the covariate imbalance between the treated and control groups before weighting (\textit{i.e.}, due to chance randomization alone) did not exceed 0.15, marked by the vertical dashed line. In most sites, weighting brought the covariate imbalance between the treated group and the target population within this range.}
    \label{fig:imbalance}
\end{figure}

Figure \ref{fig:imbalance} shows the covariate imbalance in each site before and after weighting with $\lambda = 0.03$. These weights reduce covariate imbalance across the board, with particularly reduced imbalance relative to the target distribution.

\begin{figure}
    \centering
    \includegraphics[width = 0.66\linewidth]{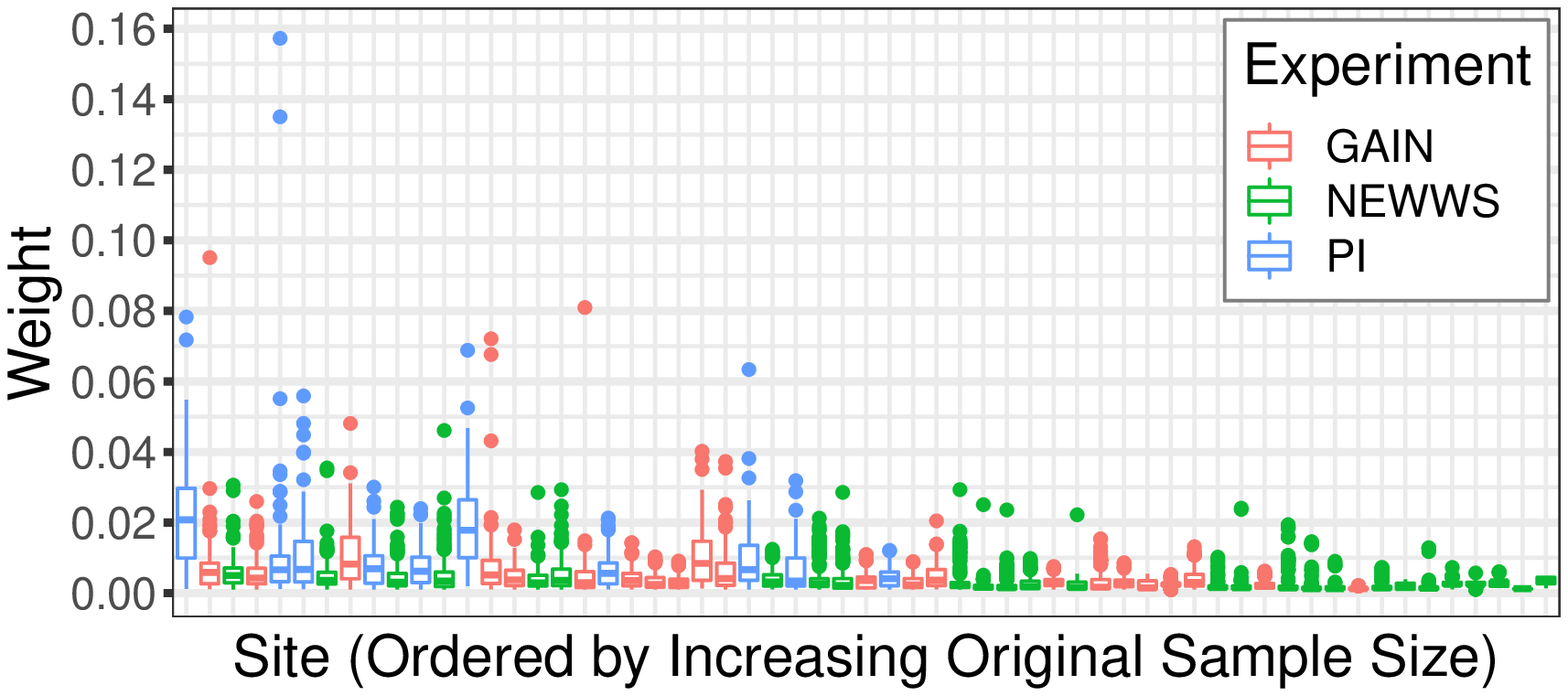}
    \includegraphics[width = 0.33\linewidth]{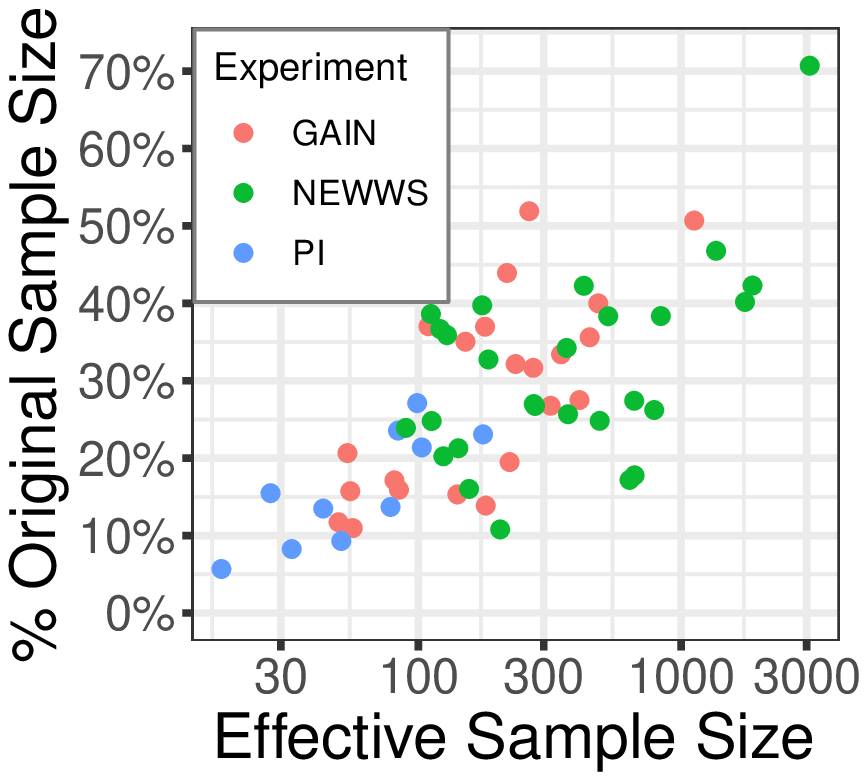}
    \caption{Distribution of non-zero weights in each site (left) and effective sample size after weighting, plotted also as a percentage of the original unweighted sample size (right).}
    \label{fig:sample_size_and_weights}
\end{figure}

To better visualize the cost of transporting the site ATE estimates, Figure \ref{fig:sample_size_and_weights} shows boxplots of the distribution of weights within each site, with weights less than $0.1\%$ excluded. The distribution of weights in some sites is heavily skewed so that a few units are given outsize weight. This lowers those sites' effective sample sizes and increases the standard errors of their transported ATE estimates. Figure \ref{fig:sample_size_and_weights} also plots the effective sample size of each site as a percentage of the site's original sample size. In most sites, the reduction in sample size was substantial but perhaps not debilitating. The median sample size before weighting was 729, while the median after weighting was 181. The smallest effective sample size in a site was 18. Overall, PI sites had more skewed distributions of weights and lower effective sample sizes than NEWWS and GAIN sites.

\begin{figure}
    \centering
    \includegraphics[width = 0.99\linewidth]{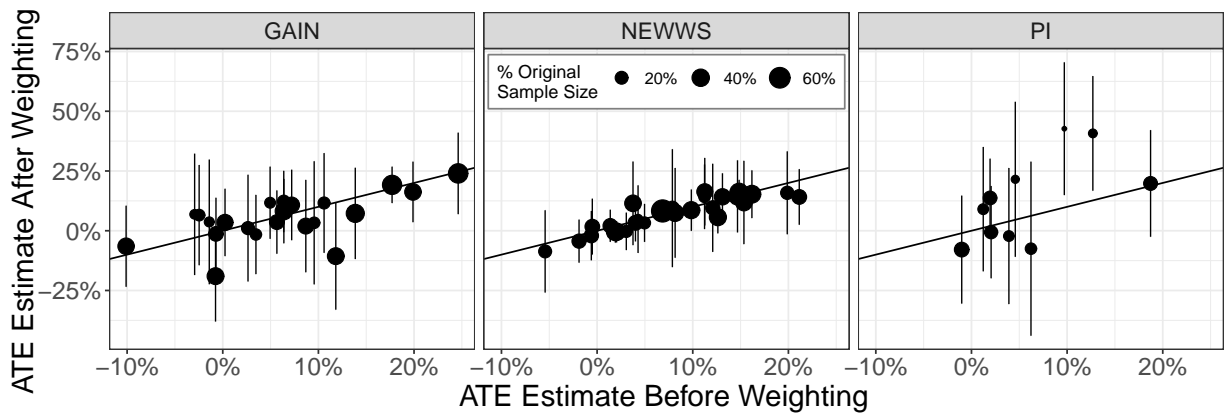}
    \caption{Point estimate and 95\% confidence interval for the transported ATE, plotted against the untransported ATE estimates. Points are sized by percent effective sample size after weighting.}
    \label{fig:all_ates_emp}
\end{figure}

Based on these tradeoffs, we choose to optimize the weights in (\ref{eq:primal_linear}) with $\lambda = 0.03$. Figure \ref{fig:all_ates_emp} shows the resulting transported ATE point and 95\% confidence interval estimate in each site, plotted against the untransported ATE estimates. Transportation substantially changed some sites' ATE estimates, though most sites saw little change.
Nor did transportation substantially change the overall variability in site ATEs.
Our $Q$-statistic analysis from Section \ref{sec:multi} estimates the standard deviation of the true untransported ATEs to be $0.060$ with a 95\% confidence interval of $(0.048, 0.075)$.
The corresponding estimate for the true transported ATEs is $0.057$ with a 95\% confidence interval of $(0.036, 0.083)$.
Using our pseudo-$R^2$ measure, we estimate that observed unit-level covariate heterogeneity explains less than 10\% of the variability in site ATEs.
So we conclude that treatment effect heterogeneity in these experiments is primarily driven by other factors such as site-level features. This is consistent with others' findings \citep[e.g.,][]{bloom2003}.

\begin{figure}
    \centering
    \includegraphics[width = 0.99\linewidth]{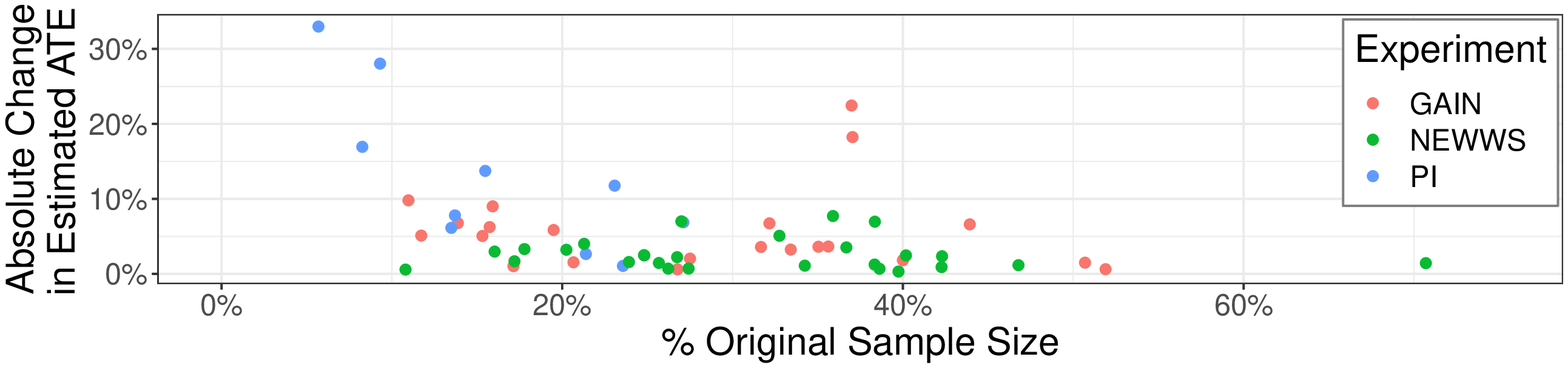}
    \caption{The absolute change in each site's estimated ATE after weighting, plotted against the effective sample size as a percentage of original sample size.}
    \label{fig:ates_v_samp}
\end{figure}

\begin{figure}
    \centering
    \includegraphics[width = 0.8\linewidth]{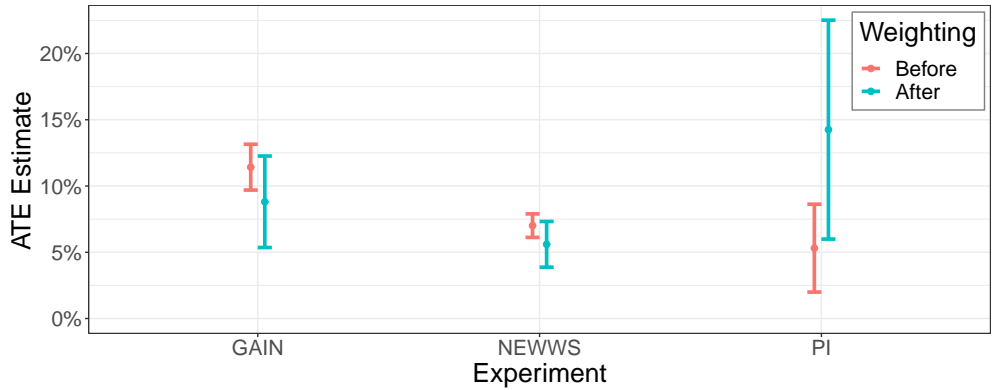}
    \caption{Point estimate and 95\% confidence interval for the ATE in each experiment before and after transportation. Transporting all of the sites to the same target distribution changed our conclusions about the relative effectiveness of treatment across the three experiments.}
    \label{fig:ates_vs_experiment}
\end{figure}

Our analysis also affects our conclusions about how effective treatment is in different types of sites.
Consider the relationship between the treatment's effect and the experiment in which the treatment was administered.
Figure \ref{fig:ates_vs_experiment} plots the estimated overall ATE in each experiment before and after transportation.
Before transportation, treatment appeared most effective in the GAIN experiment and least effective in the PI experiment.
After transporting the sites to the same target distribution of unit-level covariates, we find the NEWWS experiment least effective.
We also see a remarkable increase in the ATE estimate and associated uncertainty for the PI experiment; clearly the non-representative nature of the units in this experiment vs the overall population (note the low percentages of original sample size for the sites in Figure~\ref{fig:all_ates_emp}) make comparison beyond individual differences difficult.
That being said, its original poor average performance is quite possibly due to more difficult to treat participants.

\section{Discussion}
\label{sec:conc}

It is important for researchers and policymakers to understand how treatment effects vary across contexts. This is especially true in multisite trials, which afford the opportunity to observe treatment in a variety of settings. 
However, variation in treatment effects can reflect variation in both the population of individuals \emph{and} different contextual and site-level characteristics.
Building on the literature on treatment effect generalization and transportation, we propose to estimate the sites' ATEs if they all were to have the same population of units. We estimate these transported site-level treatment effects via an approximate balancing weights procedure and find that this procedure outperforms more traditional IPW. Applying it to the welfare-to-work experiments, we find that heterogeneity in the sites' populations accounts for less than 10\% of the overall heterogeneity in treatment effects across sites.

There are several avenues for future work. First, while we choose to re-weight to the overall population in our welfare-to-work analysis, other target populations might lead to substantively different conclusions. If there are important and substantial interactions between unit-level and site-level attributes in determining treatment effects, then we may wish to use several target populations and characterize the variation across the different targets. An important area of future work is how to characterize and present this variation in an interpretable and useful way.

Second, we assess site-level heterogeneity by using a common target for all sites. This makes the different sites more directly comparable. But we could ask a different question instead: For any given site $j$, how would the ATE in the site change if the site had the $k^{\text{th}}$ site's distribution of observed unit-level covariates, for $k \in \{1, \ldots, J\}$? This question can be answered by transporting the $j^{\text{th}}$ site's ATE estimate to each site---\textit{i.e.}, by integrating the $j^{\text{th}}$ site's CATE function over each site's observed covariate distribution. This analysis is conceptually similar to ``internal benchmarking'' in the criminology literature \citep{ridgeway2014}.

Third, we propose to construct weights by decomposing the estimation error into (1) error due to imbalance in baseline potential outcomes between the treated and control groups and (2) error due to imbalance in the CATE function between the treated group and the target population. But other decompositions are possible and can lead to different forms of the optimization problem \eqref{eq:primal} that may have different computational and statistical performance.

Finally, throughout this work, we have considered the unit-level covariates of interest to be known and fixed \emph{a priori}. However, different sets of covariates will lead to different CATE functions and different transported site treatment effects. Deciding \emph{which} covariates are of interest will be important to future work; see, e.g. \citet{Egami2019}.

%%% REFERENCES
\clearpage
\singlespacing
\bibliography{bibliography}
\bibliographystyle{chicago}

\end{document}